\documentclass[12pt]{article}
\usepackage{amsfonts}
\usepackage{amssymb}
\usepackage{hyperref}
\begin{document}

\setcounter{footnote}{1}

\begin{titlepage}

\begin{flushright}
LBNL-48641\\
UCB-PTH-01/27\\
\end{flushright}

\vskip 2cm
\begin{center}
{\large \bf Wilson Lines and Symmetry Breaking on Orbifolds}

\vskip 1.2cm 

Lawrence J. Hall, 
Hitoshi Murayama and 
Yasunori Nomura
\vskip 0.4cm

{\it Department of Physics\\
University of California, Berkeley, California 94720}\\
and\\
{\it Theoretical Physics Group\\
Ernest Orlando Lawrence Berkeley National Laboratory, MS 50A-5101\\
University of California, Berkeley, California 94720}\\
\vskip 2cm
\abstract{Gauge symmetry breaking by boundary conditions on a manifold 
is known to be equivalent to Wilson-line breaking through a background 
gauge field, and is therefore spontaneous.  These equivalent pictures 
are related by a non-periodic gauge transformation.  However, we find 
that boundary condition gauge symmetry breaking on orbifolds is explicit; 
there is no gauge where all the breaking can be attributed to a 
background gauge field.  In the case of a five-dimensional $SU(5)$ grand 
unified theory on $S^1/Z_2$, the vacuum with gauge symmetry broken to 
$SU(3) \times SU(2) \times U(1)$ and that with $SU(5)$ preserved 
are completely disconnected: there is no physical process which causes 
tunneling between the two.  This allows a certain localized explicit 
breaking of $SU(5)$ on one of the orbifold fixed points in the theory 
with $SU(5)$ breaking.  Split multiplets on this fixed point are shown 
not to induce violations of unitarity in scattering amplitudes.}

\end{center}
\end{titlepage}

\setcounter{footnote}{0}


\section{Introduction}

Symmetry breaking has been one of the most important themes in
particle physics, field theory, model building, and also in many other
areas of physics.  In most models of symmetry breaking in field theory, 
there is a scalar field, with a non-trivial quantum number under the 
symmetry, acquiring a vacuum expectation value, which is the order 
parameter.  In the case that the symmetry is local, {\it i.e.}\/ 
a gauge symmetry, the Higgs mechanism operates and the longitudinal 
components of the gauge fields are generated by ``eating'' the fields 
responsible for symmetry breaking.  As is well-known, an explicit 
breaking of gauge symmetry, such as adding mass terms to gauge bosons, 
leads to an inconsistent theory, breaking unitarity at high energies. 
However, a spontaneous symmetry breaking does not modify the high-energy 
behavior of the theory and the theory remains consistent despite massive 
gauge bosons \cite{LlewellynSmith:1973ey}.

Recently, there has been a strong interest in studying symmetry 
breaking phenomena using symmetry-violating boundary conditions in 
extra dimensions \cite{Antoniadis:1993fh, Pomarol:1998sd, 
Antoniadis:1999sd, Barbieri:2001vh, Arkani-Hamed:2001mi, Delgado:2001si, 
Kawamura:2000ev, Altarelli:2001qj, Hall:2001pg, Nomura:2001mf, 
Hebecker:2001wq, Barbieri:2001yz, Barbieri:2001dm, Hebecker:2001jb, 
Bagger:2001qi, Masiero:2001im}.  Grand-unified symmetry 
and supersymmetry can be broken by gauge non-invariant and 
non-supersymmetric boundary conditions, respectively, via the
Scherk--Schwarz mechanism \cite{Scherk:1979ta}.  Natural questions are:
what is the order parameter? What degrees of freedom are ``eaten''
by the gauge field? Is the breaking spontaneous or explicit?
In the case of supersymmetry breaking, Marti and Pomarol suggested 
that the order parameter in the Scherk--Schwarz mechanism is the 
$F$-component of the radion field, the fluctuation of the size of 
the extra dimensions, while its fermionic component supplies the 
longitudinal components for the gravitino \cite{Marti:2001iw}.  
The questions we ask here are concerned with gauge symmetry breaking.

On multiply-connected manifolds, it is well-known that gauge 
symmetries may be broken by the presence of background gauge fields.
This phenomenon has been known as the Hosotani mechanism, or Wilson-line
symmetry breaking \cite{Hosotani:1983xw, Candelas:1985en}.  
We review this mechanism in section \ref{sec:hol-man}, as we wish 
to explore its relation to symmetry breaking by boundary conditions.
This relationship is also discussed in \cite{Hebecker:2001jb}. When 
one classifies possible vacua for gauge fields on multiply-connected
compact manifolds $M$, one finds that it is equivalent to the
classification of flat bundles, which in turn can be classified 
by the ``representations'' (homomorphisms) of the fundamental group 
$\pi_1(M)$ into the gauge group.  The associated gauge group element 
for each cycle is the holonomy.  In section \ref{sec:holon} 
we discuss how this idea naturally extends to orbifolds, where the 
concept of the fundamental group is ill-defined. Some of these 
concepts were introduced in the context of studying low energy limits 
of string theory \cite{Dixon:jw}. In section \ref{sec:flavor} we discuss 
how discrete flavor orbifold symmetries of an effective field theory 
may arise from a more fundamental gauge theory at higher energies.

In the five-dimensional $SU(5)$ theory considered in 
\cite{Kawamura:2000ev, Hall:2001pg}, the origin of gauge symmetry 
breaking is viewed as a boundary condition on the orbifold $S^1/Z_2$. 
In section \ref{sec:SU5} we study alternative formulations of this 
theory involving non-trivial background gauge fields, and find that 
the boundary conditions always involve $SU(5)$ breaking. Hence, in 
contrast to the case of a manifold, the gauge symmetry breaking in 
the effective field theory on an orbifold is explicit, allowing certain 
special explicit breaking on fixed points of the orbifold. In section 
\ref{sec:fixed} we study this local explicit breaking of the gauge 
symmetry on the fixed points.  Non-unified multiplets and their 
interactions, respecting only the unbroken symmetries on the brane, 
are allowed, while the physics in the bulk is still constrained by 
the full gauge invariance \cite{Hall:2001pg}.  We check that this 
explicit breaking does not lead to additional violations of unitarity 
in certain scattering amplitudes. In section \ref{sec:concl} we 
present our conclusions.

\section{Holonomies on Manifolds}
\label{sec:hol-man}

Let us first review Wilson Lines on compact manifolds.  Wilson lines 
have been used to break gauge symmetries when extra dimensions are 
compactified \cite{Candelas:1985en, Breit:1985ud, Derendinger:1986cv}.

Consider a gauge theory on a compact manifold $M$ with the standard
Hamiltonian
\begin{equation}
    H = \int dx \frac{1}{2} \left[(\vec{E}^{a})^{2} + 
    (\vec{B}^{a})^{2}\right].
\end{equation}
The question is what the ground-state configurations are for the 
gauge field.  

In the temporal (or Weyl) gauge where $A_{0}^{a} = 0$, the Hamiltonian 
simplifies to
\begin{equation}
    H = \int dx \frac{1}{2} \left[(\dot{\vec{A}^{a}})^{2} + 
    (\vec{B}^{a})^{2}\right].
\end{equation}
For time-independent configurations, the first term simply vanishes.  
Because the Hamiltonian is positive semi-definite, the ground-state 
configuration is given by the equation
\begin{equation}
    \vec{B}^{a} = 0.
\end{equation}
On a simply-connected space (a connected space with no non-contractible
cycles), the solution to this equation is simply that the gauge field
is a pure-gauge configuration.

However, the situation is much more interesting on
non simply-connected manifolds.  Let us take a plane with the origin
removed $M={\mathbb R}^{2}\backslash\{0\}$ and $U(1)$ gauge group as 
an example.  The condition that the field strength vanishes $B = 0$
allows the vector potential
\begin{equation}
    (A_{x}, A_{y}) = \frac{\Phi}{2\pi} \vec{\nabla} \arctan \frac{y}{x}
    = \frac{\Phi}{2\pi} (-\frac{y}{r^{2}}, \frac{x}{r^{2}}).
\end{equation}
Because the vector potential is a gradient, {\it i.e.}\/ pure gauge, 
the field strength vanishes identically.  There is, however, the 
Aharonov--Bohm phase
\begin{equation}
    \exp \left( i \frac{e}{\hbar c} \oint \vec{A} \cdot d\vec{x}\right)
    = \exp \left( i \frac{e\Phi}{\hbar c} \right).
\end{equation}
Unless $\Phi \neq 2\pi n \hbar c/e$ for $n \in {\mathbb Z}$, the gauge
field has a physical significance.  Unlike the Aharonov--Bohm effect,
however, where one can imagine looking at the origin and measure the
magnetic flux $\Phi$ there, the origin is {\it outside}\/ the
manifold in this example, and the magnetic flux is ill-defined.  
In fact, any magnetic fluxes which differ by an integer multiple of
$2\pi \hbar c/e$ give identical physics on $M$.  This is because one 
can perform a gauge transformation by
\begin{equation}
    U = \exp \left( i n \arctan\frac{y}{x}\right),
\end{equation}
which is single-valued on $M$.  We find
\begin{equation}
    \vec{A} \rightarrow \vec{A} - i \frac{\hbar c}{e} U^{-1} 
    \vec{\nabla} U = \vec{A} + n \frac{\hbar c}{e}
    (-\frac{y}{r^{2}}, \frac{x}{r^{2}}).
\end{equation}
Therefore, $\Phi$ and $\Phi + 2\pi n\hbar c/e$ are gauge-equivalent. 
From this point on, we set $\hbar=c=1$ and the coupling constant $e$
absorbed in the definition of the gauge field.

It is sometimes more convenient to try to ``gauge away'' the gauge 
field as much as possible.  For the above example, we can try
\begin{equation}
    U = \exp \left( i \frac{\Phi}{2\pi} \arctan\frac{y}{x}\right).
\label{eq:U}
\end{equation}
The problem with this gauge transformation is that it is not
single-valued.  If we start from the positive $x$-axis and go around
the origin counter-clockwise, the phase of $U$ in Eq.~(\ref{eq:U})
monotonically increases, becoming $\Phi$ when we come back to the 
positive $x$-axis from below.  This is, in general, not an integer 
multiple of $2\pi$ and hence $U$ is discontinuous.  We need to put 
a ``cut'' along the positive $x$-axis.  The gauge field is now 
identically zero all around, but when one crosses the cut, we need 
a gauge transformation by $\exp (i \Phi)$ (``transition function'').  
One can view this as a boundary condition for fields when going 
around the origin and do not have to refer to gauge field background 
anymore.  Therefore, ground-state gauge field configurations can be 
classified by assigning Aharonov--Bohm phases or transition functions 
for non-contractible cycles without writing down explicit forms of 
the gauge fields \cite{Hosotani:1983xw, Candelas:1985en}.

Let us discuss a simple example of $M = S^{1}$ parameterized by $y \in
[0, 2\pi R]$ and $G=SU(2)$.  We take the gauge field to be along 
the 3rd isospin direction
\begin{equation}
    A(y) = A^{3}(y) \frac{1}{2} \tau^{3} = A^{3}(y) \frac{1}{2} 
    \left( \begin{array}{cc} 1 & 0\\ 0 & -1 \end{array} \right),
\end{equation}
where $A$ is the spatial component of the gauge field. Without 
loss of generality, one can make $A^{3}$ constant by use of gauge 
transformations.  Moreover, the constant $A^{3}$ can be further 
gauge-transformed using the single-valued gauge transformation function
\begin{equation}
    U(y) = \exp \left( i \tau^{3} \frac{y}{R} \right),
    \label{eq:large}
\end{equation}
as
\begin{equation}
    A^{3} \rightarrow A^{3} + \frac{2}{R}.
\end{equation}
Therefore $A^{3} R$ is defined mod 2.  The holonomy, $T$, is given by
\begin{equation}
  T = \exp \left( i \oint A^{3} \frac{\tau^{3}}{2} dy \right)
    = \left( \begin{array}{cc}
        e^{\pi i A^{3}R} & \\
        & e^{- \pi i A^{3}R}
        \end{array} \right).
\end{equation}
The spectrum of particles is affected by the presence of the constant
gauge field, or equivalently, a non-trivial transition function.  If
we use the language of the constant gauge field, the Kaluza--Klein
(KK) modes of a Klein--Gordon field $\phi$ of isospin $I_{3} = 1/2$ 
reads as
\begin{equation}
    -\left(\partial_{y} - i \frac{1}{2} A^{3} \tau^{3} \right)^{2} \phi 
    = m_{KK}^{2} \phi .
\end{equation}
$\phi$ is periodic: $\phi(y+2\pi R) = \phi(y)$.  Then the $n$-th mode 
has the form $\phi_{n}(y) = e^{i n y/R}$, and
\begin{equation}
    -\left(\partial_{y} - i \frac{1}{2} A^{3} \tau^{3} \right)^{2} \phi_{n} 
    = \left( \frac{n}{R} - \frac{1}{2} A^{3} \tau^{3} \right)^{2} \phi_{n}.
\end{equation}
Therefore the KK mass is $m_{n} = \left|\frac{n}{R} -
\frac{1}{2} A^{3} \tau^{3} \right|$.  The gauge transformation
Eq.~(\ref{eq:large}) would lead to a relabeling of states by $n
\rightarrow n\pm 1$, but the physical spectrum remains unchanged as it
should.  If instead we had gauged away the gauge field by
\begin{equation}
    U(y) = \exp \left(- \frac{i}{2} A^{3} \tau^{3} y\right),
\label{eq:non-peri}
\end{equation}
the mode equation is simply
\begin{equation}
    -\partial_{y}^{2} \phi = m_{KK}^{2} \phi,
\end{equation}
while the boundary condition is changed to
\begin{equation}
    \phi(2\pi R) = e^{-\pi i A^{3} \tau^{3} R} \phi(0).
\end{equation}
The modes are given by
\begin{equation}
    \phi_{n}(y) = \exp \left[ i\left(\frac{n}{R} - 
    \frac{1}{2} A^{3} \tau^{3} \right) y \right],
\end{equation}
giving the same spectra as obtained using a constant gauge 
field and periodic $\phi$.  We find that the non-periodic gauge 
transformation in Eq.~(\ref{eq:non-peri}) can interpolate between 
two different (but equivalent) pictures: one in which the boundary 
condition for $\phi$ is trivial and gauge breaking comes entirely 
from the non-vanishing background gauge field (Wilson lines), and 
the other in which the expectation value for the gauge field is 
vanishing and the gauge breaking comes from the non-trivial boundary 
condition of $\phi$ (and the gauge fields).  Therefore, we conclude 
that the boundary condition gauge breaking is spontaneous, since we 
can always go to a basis where the boundary conditions for the fields 
are trivial and the gauge breaking arises solely from the non-zero 
vacuum expectation value of the gauge field, $\langle A \rangle$. 
We will see later that this is not necessarily the case for gauge 
symmetry breaking on orbifolds, leading to a remarkable feature for 
the orbifold gauge breaking: the presence of local defects of the 
gauge symmetry on the fixed points.

The KK spectrum of the gauge field is also affected by the
holonomy.  Because $A_{\mu}^{3}$ commutes with $A^{3}$ itself, the 
spectrum of the unbroken $U(1)$ is simple.  Here, $\mu$ represents 
the index of uncompactified spacetime dimensions. The mode functions 
are $A_{\mu,n}^{3} (y) = \cos n y/R$ and $\sin ny/R$, and the mass 
spectrum is $n/R$ with a single zero mode: the gauge field of the 
unbroken $U(1)$.  On the other hand, the $W^{\pm}$-bosons become 
massive.  Their isospins are $I_{3} = \pm 1$, and the same analysis 
for the Klein--Gordon field leads to the KK $W$-spectrum of 
$m_{n} = \left| \frac{n}{R}\mp A^{3} \right|$.

In general, vanishing field-strength configurations (or ``flat
connections'') on non-simply connected manifolds can be classified 
in terms of Aharonov--Bohm phases (``holonomies'') for each
non-contractible cycles.  ``Magnetic fluxes'' are ``outside'' the
manifold and have no physical meaning; however the holonomies do. 
The cycles can be continuously deformed without changing the phase,
because the difference in the phase can be expressed as an area
integral of the magnetic field using Stokes' theorem which vanishes
because of the vanishing field strength.  Therefore a holonomy is
associated with a class of non-contractible cycles that can be
continuously deformed among each other.  For each class of
non-contractible cycles $c_{i}$, we associate a gauge group
element $g_{i} \in G$ by 
\begin{equation}
    g_{i} = {\cal P} \exp \left( i \oint_{y_{0}, c_{i}} A(y) dy \right),
\end{equation}
where ${\cal P}$ denotes the path-ordered product.  Note that the 
cycles are defined with respect to an origin $y_{0}$ (even though 
the group itself does not depend on the choice of the origin) and 
the associated holonomies are also defined with respect to the origin. 
Under gauge transformations, the elements transform as
\begin{equation}
    g_{i} \rightarrow U(y_{0}) g_{i} U(y_{0})^{-1}.
\end{equation}
Just like in the Abelian example at the beginning of the section,
$g_{i}$ does not change under the continuous deformation of the cycle
$c_{i}$ because the field strength vanishes.  When one goes around
non-contractible cycles successively, in some cases it matters in
which order one goes around different cycles.  Successive loops can 
be viewed as a product of two cycles, defining a group among
non-contractible cycles: the fundamental group or the first homotopy
group $\pi_{1} (M)$.  In the above example of a plane with the origin
removed, $\pi_{1}(M) = {\mathbb Z}$ because a non-contractible cycle
is just labeled by the number of times it goes around the origin. 
For more complicated manifolds, such as a genus-two Riemann surface
(two-hole doughnut), the fundamental group is non-Abelian.

The ground-state configurations of the gauge field are classified 
by the gauge-inequivalent classes of holonomies.  In other words,
\begin{equation}
    \{ \mbox{inequivalent ground state configurations} \}
    = \textrm{Hom} (\pi_{1} (M) \rightarrow G)/G,
    \label{eq:def1}
\end{equation}
where the r.h.s. is the set of all homomorphisms (maps that respect 
multiplications of elements) from the fundamental group $\pi_{1}(M)$ 
to the gauge group $G$, modded out by the gauge equivalence $G$.  
As we explained before, these holonomies can be viewed either as 
a non-trivial background gauge field (Wilson loop associated with 
a non-contractible cycle) or as boundary conditions imposed on the 
fields when we move along the non-contractible cycle.

When the manifold $M$ is defined as a coset $M_{0}/H$ of a
simply-connected manifold $M_{0}$ by a discrete symmetry group $H$,
the fundamental group is simply $H$ itself.  This is the consequence
of the homotopy exact sequence
\begin{equation}
    \cdots \rightarrow \pi_{n} (M_{0}) \rightarrow 
    \pi_{n} (M_{0}/H) \rightarrow \pi_{n-1} (H) \rightarrow 
    \pi_{n-1} (M_{0}) \rightarrow \cdots.
\end{equation}
Because of the assumption that $M_{0}$ is simply-connected,
$\pi_{1}(M_{0}) = \pi_{0}(M_{0}) = 0$,\footnote{The zeroth homotopy
group $\pi_{0}$ refers to the disconnected parts of the space.} 
we obtain the exact sequence
\begin{equation}
    0 = \pi_{1}(M_{0}) \rightarrow \pi_{1} (M_{0}/H) \rightarrow 
    \pi_{0} (H) \rightarrow \pi_{0} (M_{0}) = 0.
\end{equation}
For this to be an exact sequence, we find
\begin{equation}
    \pi_{1} (M_{0}/H) = \pi_{0} (H) = H,
\end{equation}
where the last equality holds because we assumed that $H$ is a 
discrete group.  Then the space of ground-state configurations is
\begin{equation}
    \{ \mbox{inequivalent ground state configurations} \}
    = \textrm{Hom} (H \rightarrow G)/G,
\label{eq:def2}
\end{equation}
or equivalently, a set of all possible ``representations'' of 
the discrete group $H$ in $G$ modulo gauge equivalences.

It is instructive to look at an example where $H$ is a discrete group. 
For instance, consider $M_{0}=S^{2}$, and $H=Z_{2}$ identifies the
anti-podal points of the sphere, so that $M = S^{2}/Z_{2} = {\mathbb 
R}P^{2}$.\footnote{This manifold is unorientable, but it does not
matter for the purpose of this discussion.} On $S^{2}$, scalar fields
(or gauge fields in remaining uncompactified dimensions) are expanded
in terms of spherical harmonics $Y_{l}^{m}$ with $m^{2} = l(l+1)$. 
Note that under $Z_{2}$, $Y_{l}^{m} \rightarrow (-1)^{l} Y_{l}^{m}$. 
On ${\mathbb R}P^{2}$, one can have the holonomy $T = {\rm diag}(-1,
-1, +1)$ in the $SU(3)$ gauge group.  The main difference from the
$S^{1}$ case discussed earlier in this section is that the choice for 
$T$ is discrete.  Then the unbroken $SU(2) \times U(1)$ gauge bosons have
even $T$, and hence only even $l$ are allowed, while the broken gauge
bosons have odd $T$, and only odd $l$ are allowed.  This does not
allow the broken gauge bosons to be massless as expected.

\section{Holonomies on Orbifolds}
\label{sec:holon}

An orbifold is defined as a coset of a manifold by a discrete 
symmetry.  If the symmetry does not act freely, {\it i.e.}\/ there 
are fixed points that do not move under the symmetry, the coset space 
is singular at the fixed points and is not a manifold.  It is an 
orbifold instead.  Because it is not a manifold, the conventional 
definition of the homotopy groups fails.  Therefore, we cannot 
classify the ground-state configurations of the gauge field using 
Eq.~(\ref{eq:def1}).  However, Eq.~(\ref{eq:def2}) is still 
well-defined because the very definition of an orbifold is in terms 
of a coset space. 

For example, consider the $S^{1}/{\mathbb Z}_{2}$ orbifold.  
This is a circle $y \in [0, 2\pi R]$ modded out by the ``parity'' 
${\mathbb Z}_{2}$ symmetry $P: y \rightarrow -y$.  Of course 
$P^{2} = 1$.  We cannot apply Eq.~(\ref{eq:def2}) because $S^{1}$ 
is not simply connected.  However, $S^{1}$ itself can be viewed as 
a coset space ${\mathbb R}/{\mathbb Z}$, where ${\mathbb R}$ is the 
real axis for $y$ and ${\mathbb Z}$ is generated by the translation 
$T: y \rightarrow y + 2\pi R$.  Then the orbifold $S^{1}/{\mathbb Z}_{2} 
= {\mathbb R} / ({\mathbb Z} \ltimes {\mathbb Z}_{2})$. 
The translation $T$ and the parity do not commute,
\begin{equation}
    P T P = T^{-1},
    \label{eq:PTPT}
\end{equation}
and hence the semi-direct product $H={\mathbb Z} \ltimes 
{\mathbb Z}_{2}$.\footnote{For discussions on the constraints on the 
possible holonomies in orbifolds, see \cite{Dixon:jw}.} 
Because ${\mathbb R}$ is simply-connected, we can now use 
Eq.~(\ref{eq:def2}) to find the space of ground-state configurations: 
all representations of ${\mathbb Z} \ltimes {\mathbb Z}_{2}$ in the 
gauge group $G$.  Note that Eq.~(\ref{eq:PTPT}), together with 
$P^{2}=1$, defines the dihedral group $D_{n}$ if $T^{n}=1$. Therefore 
the orbifold group here is the $n\rightarrow \infty$ limit of the 
dihedral groups.  The holonomy $T$ in this example can be represented 
as a Wilson line in exactly the same fashion as on manifolds,
\begin{equation}
    T = {\cal P} \exp \left( i \oint_{0}^{2\pi R} A(y) dy \right),
\end{equation}
where ${\cal P}$ denotes the path-ordered product.  However, the 
unitary matrix representing $P$ cannot be expressed in terms of 
a gauge-field background.  

In general, elements for freely acting generators of $H$ can be
expressed in terms of background gauge fields, while those for
non-freely acting generators of $H$ (those with fixed points) 
cannot be.  One may still view the whole theory to be $G$ invariant, 
with the non-trivial holonomy as the ``expectation values'' of the 
gauge fields.  However, although the holonomies for non-freely acting 
generators of $H$ are a part of the classification of the symmetry 
breaking as given in Eq.~(\ref{eq:def2}), they cannot be expressed 
in terms of the background gauge fields. As a consequence, if one 
considers general gauge symmetry breaking on orbifolds, one cannot 
necessarily go to the basis where the boundary conditions for the 
fields are all trivial and the gauge breaking entirely comes from 
the expectation value of the background gauge field. This implies 
that the boundary condition gauge breaking on orbifolds is not 
completely spontaneous, and some of the gauge breaking effects must 
come intrinsically from the boundary conditions.  This fact allows 
special point-like symmetry-breaking defects on the orbifold fixed 
points as shown in section \ref{sec:SU5} for an explicit example.

When $G=U(1)$, both $T$ and $P$ are represented as phases, and hence 
they commute.  Even when $G$ is non-Abelian, $T$ and $P$ may commute 
for certain representations and the following discussions still apply. 
Then $T^{2}=1$ from Eq.~(\ref{eq:PTPT}) and only possibilities are 
$T = \pm 1$.  Because $P^{2}=1$, we also find $P = \pm 1$.  Let us 
check this result based on a formal argument explicitly using $G=SU(2)$ 
as an example.  We take the gauge field to be along the 3rd isospin 
direction as in the previous section,
\begin{equation}
    A(y) = A^{3}(y) \frac{1}{2} 
    \left( \begin{array}{cc} 1 & 0\\ 0 & -1 \end{array} \right).
\end{equation}
Again without loss of generality, one can make $A^{3}$ constant by use 
of gauge transformations. Moreover, the constant $A^{3}$ can be further 
gauge-transformed using Eq.~(\ref{eq:large}) as
\begin{equation}
    A^{3} \rightarrow A^{3} + \frac{2}{R}.
\label{eq:A3-trans}
\end{equation}
Therefore $A^{3} R$ is defined mod 2.  Under the parity, the gauge 
field $A(y)$ transforms as $A(y) \rightarrow -A(-y)$, because the 
covariant derivative $\partial_{y} - i A(y)$ must transform covariantly. 
(Here and hereafter, we normalize the gauge field such that the kinetic 
term has the coefficient $1/g^{2}$ while the covariant derivative does 
not have the coupling constant.)  This fixes $P$ on the gauge field 
uniquely to be $P = -1$.  Naively, it appears that the gauge field 
must vanish identically because of the parity invariance.  However, 
it is possible that the gauge field transforms to its gauge-equivalent 
configuration under parity, and hence 
\begin{equation}
    A^{3} = \frac{1}{R} \label{eq:A3=1},
\label{eq:vac-conf}
\end{equation}
is perfectly allowed.  Under parity $y \rightarrow -y$ it transforms 
to $-1/R$, but is gauge equivalent to $+1/R$ by the use of the gauge 
transformation Eq.~(\ref{eq:large}).  In other words, if we redefine 
the parity as the combined transformation of $y \rightarrow -y$ 
and the gauge transformation Eq.~(\ref{eq:large}), the vacuum 
configuration Eq.~(\ref{eq:vac-conf}) is invariant.  Then the holonomy 
for going around the circle is
\begin{equation}
    T = \exp \left( i \oint A^{3} \frac{\tau^{3}}{2} dy \right)
    = -1 .
\end{equation}
Therefore Eq.~(\ref{eq:A3=1}) is a consistent ground-state configuration 
of the gauge field which corresponds to the possibility found from 
the formal argument.  An important point here is that the ``parity'' 
involves a non-trivial operation on the gauge space.  Namely, the 
``gauge twist'' lies not only in the expectation value for $A^3$ but 
also in the definition of the parity transformation. We note that this
choice of $T$ is in the center of the gauge group, and hence does not 
break the gauge invariance.  Another way to see it is that $T = -1$ on 
the fundamental representation (doublet) is trivial, $T=1$, on integer 
isospin representations.  The $W$-bosons with the spectrum $m_{n} = 
\left| \frac{n}{R}\mp A^{3}\right|$ indeed have zero modes.  The only 
effect then is on half-odd isospin fields that can be viewed simply as 
anti-periodic boundary conditions for them under the shift $T$. 

The fact that the gauge group is not broken with Eq.~(\ref{eq:A3=1})
is a special situation with the $SU(2)$ gauge group where the only 
non-trivial choice for the holonomy was in the center.  To break 
the $SU(2)$ gauge invariance, we have to consider the case with 
$A^{3} = 1/(2R)$.  This is possible if the theory possesses the 
discrete gauge symmetry under which all the half-odd isospin fields 
$\phi$ transform as $\phi \rightarrow \pm i\phi$.  Then, combined with 
the non-trivial holonomy in the flavor space, $\phi \rightarrow \pm i\phi$, 
we can break the $SU(2)$ gauge symmetry.  A similar situation arises 
when one tries to achieve, {\it e.g.}\/, $SU(2n) \rightarrow SU(n) 
\times SU(n) \times U(1)$ breaking.  This issue is discussed in the 
next section.

An example which breaks the gauge invariance by Eq.~(\ref{eq:A3=1})
is $SU(5)$ with $T = {\rm diag}(1, 1, 1, -1, -1)$, which has 
interesting phenomenological applications.  We will see this example 
in more detail in section \ref{sec:SU5}.  We also explicitly see there 
that all gauge breaking cannot be attributed to the expectation value 
for the gauge field in certain orbifold theories, leading to an explicit 
symmetry breaking defect on the fixed point of the orbifold.  A somewhat 
simpler example is the $SU(3)$ gauge group with the holonomy 
$T = {\rm diag}(-1, -1, 1)$.  With the choice analogous to that in 
Eq.~(\ref{eq:A3=1}), the gauge group is indeed broken to $SU(2) \times 
U(1)$, and it is easy to work out the spectrum and verify explicitly 
that the gauge bosons for the broken generators indeed do not have 
zero modes.

Another interesting possibility is a non-commuting $P$ and $T$.  For
instance, when $G=SU(3)$ the following representation is possible, 
motivated by the connection to the dihedral groups:
\begin{equation}
    P = \left( \begin{array}{ccc} 1 & &\\ & -1 &\\ & & -1
    \end{array} \right), \qquad
    T = \left( \begin{array}{ccc} \cos\theta & \sin\theta & 0\\
    -\sin\theta & \cos\theta & 0 \\ 0 & 0 & 1 \end{array} \right).
\end{equation}
Such a possibility would be useful to reduce the rank of the gauge 
group.  Here we note that the effect of $T$ can be attributed 
completely to the expectation value for the gauge field by making an 
appropriate non-periodic gauge transformation, while that of $P$ can not. 
This means that the gauge breaking by $P$, $SU(3) \rightarrow SU(2) 
\times U(1)$, is intrinsically an orbifolding one, which for example 
leads to explicit local symmetry-breaking defects, while the breaking 
by $T$ is completely spontaneous, which can be understood as the breaking 
caused by the Higgs (Hosotani) mechanism by an expectation value for 
the gauge field.  In general, the gauge breaking caused by a continuous 
parameter in the boundary conditions can be viewed as spontaneous breaking, 
since we can always eliminate that parameter using non-periodic gauge 
transformations and attribute all the breaking effect to the expectation 
value of the gauge field, as we have explicitly seen in the manifold 
case in section \ref{sec:hol-man}.  On the other hand, the breaking 
caused by a discrete operation, such as the one in $P$ in the above 
example, is intrinsically orbifold breaking, leading to new interesting 
phenomena as we will see explicitly in section \ref{sec:fixed}.

\section{Flavor Holonomy}
\label{sec:flavor}

For matter fields, there are often additional flavor symmetries.  
Then one can assign non-trivial holonomies in the flavor space as 
well. For instance, a real scalar field with the potential $V = m^{2} 
\phi^{2} + \lambda \phi^{4}$ has a ${\mathbb Z}_{2}$ symmetry $\phi 
\rightarrow -\phi$.  Then for each non-contractible cycle, we can 
assign either even or odd ${\mathbb Z}_{2}$ element under the flavor 
symmetry, {\it i.e.}\/ periodic or anti-periodic boundary conditions.  

When can we assign holonomies for non-gauge symmetries?  We can do so 
when we can regard the flavor symmetry ``gauged'' in the following 
sense \cite{Krauss:1989zc}.  The above ${\mathbb Z}_{2}$ model can 
be promoted to a full $U(1)$ model by adding an imaginary part to the 
scalar field $\phi + i \eta$.  The $U(1)$ gauge group is broken to 
${\mathbb Z}_{2}$ by an expectation value of the field $\psi$ with 
charge $+2$.  Without loss of generality, we can take $\langle \psi 
\rangle \in {\mathbb R}$.  Then using a term $(\phi-i\eta)^{2} 
\langle\psi\rangle + (\phi+i\eta)^{2} \langle\psi^{*}\rangle = 
2(\phi^{2} - \eta^{2}) \langle \psi \rangle$, we can make $\phi$ and 
$\eta$ non-degenerate; the low-energy effective theory may well have 
only $\phi$ but not $\eta$.  On the other hand, the $U(1)$ gauge group 
is broken by the charge two field and hence allows a ``cosmic string'' 
solution where the $\psi$ field has its phase rotated by $2 \pi$ as one 
goes around a non-contractible cycle.  For a charge one field, however, 
the phase rotates by only $\pi$; this situation is equivalent to having 
the holonomy of $-1$ under the $U(1)$ gauge group.  In this sense, the 
anti-periodic boundary condition can be viewed as a non-trivial 
holonomy of the ``gauged ${\mathbb Z}_{2}$ symmetry''.  Once taken 
this view, there are non-trivial constraints on what flavor holonomies 
can exist.  For instance, the flavor symmetry must be anomaly-free 
\cite{Ibanez:1991hv}.

In some cases, the existence of the non-trivial holonomy of gauged 
discrete symmetries is essential to break gauge symmetries. (This 
corresponds to the case $P_G^2 = P_H^2 = -1$ in \cite{Barbieri:2001dm}.)
This happens, for instance, in the case $G=SU(2)$.  As we discussed 
in the previous section, if there were no discrete gauge symmetry, 
the gauge transformation function which does not change the boundary 
conditions of the fields is given by Eq.~(\ref{eq:large}).  Then, the 
only vacuum configuration allowed would be Eq.~(\ref{eq:A3=1}), since 
the gauge transformation of $A^3$ is given by Eq.~(\ref{eq:A3-trans}).
However, if the $SU(2)$ gauge theory possesses the ${\mathbb Z}_{4}$ 
discrete gauge symmetry under which all the half-odd isospin fields 
have charges $\pm 1$, then we can use the non-trivial holonomy of this 
${\mathbb Z}_{4}$ symmetry to change the situation. Specifically, 
we can consider the vacuum configuration 
\begin{equation}
    A^{3} = \frac{1}{2R}, 
\end{equation}
together with the boundary condition for half-odd isospin fields $\phi$ 
\begin{equation}
    \phi(y+2\pi R) = \pm i \phi(y),
\end{equation}
since these vacuum configuration and boundary conditions are invariant 
under the ``parity'' that is defined as the combined transformation 
of $y \rightarrow -y$ and 
\begin{equation}
    U(y) = \exp \left( i \tau^{3} \frac{y}{2R} \right).
\end{equation}
In this case, the holonomy in the gauge space is given by 
\begin{equation}
    T_G = \exp \left( i \oint A^{3} \frac{\tau^{3}}{2} dy \right)
    = i \tau^{3},
\end{equation}
and thus the $SU(2)$ gauge symmetry is broken.
On the other hand, the holonomy in the flavor space is 
\begin{equation}
    T_F = \pm i,
\end{equation}
for the half-odd isospin fields ($T_F = \pm 1$ for integer isospin fields).
Therefore, the consistency condition $T^2 = (T_G \otimes T_F)^2 = 1$ 
still holds (all the fields transform as $\pm 1$ under $T$).

\section{$SU(5)$ Breaking on $S^1/Z_2$}
\label{sec:SU5}

In this section we study the $SU(5)$ gauge breaking theory of 
\cite{Kawamura:2000ev, Hall:2001pg} in a gauge where the translation 
boundary condition is trivial.  We find that it is not possible to 
remove $SU(5)$ breaking effects from the entire boundary conditions, 
and thus show that this breaking cannot be viewed as spontaneous.
We also clarify the notion of the restricted gauge symmetry 
introduced in \cite{Hall:2001pg} in the present context.

The model we study is based on a five-dimensional theory with the gauge 
group $SU(5)$ broken by non-trivial orbifold boundary conditions.
The $SU(5)$ gauge transformation acts on various fields as
\begin{eqnarray}
    \phi(x^\mu, y) &\rightarrow& U \phi(x^\mu, y),
\\
    A_5^a T^a (x^\mu, y) &\rightarrow& 
    U A_5^a T^a (x^\mu, y) U^{-1} - i U^{-1} \partial_y U,
\end{eqnarray}
where $U = \exp(i \theta^a (x,y) T^a)$ and $A_5$ is the 5th component 
of the gauge field.  To achieve the desired breaking of the $SU(5)$ 
symmetry, we consider the vacuum configuration such that $A_5$ has 
an expectation value in the hypercharge direction.  Therefore we 
concentrate on this direction from now on.  Then, the above 
transformations can be written as 
\begin{eqnarray}
    \phi(x^\mu, y) &\rightarrow& 
    {\rm e}^{i \theta(x^\mu,y) Y_\phi} \phi(x^\mu, y),
\\
    A_5(x^\mu, y) &\rightarrow& 
    A_5(x^\mu, y) + \partial_y \theta(x^\mu,y).
\end{eqnarray}
Here, $A_5$ and $\theta$ are in the hypercharge direction and 
the hypercharges $Y_\phi$ of $\phi$ are normalized such that they 
take integer or half integer values.

As in section \ref{sec:holon}, we consider the large gauge transformation 
with $\theta(x^\mu,y)$ linear in $y$.  Then, if we require the periodicity 
of the gauge transformation under $y \rightarrow y + 2\pi R$, 
\begin{equation}
    \theta(x^\mu, y) = \frac{2n}{R} y,
\label{eq:theta-su5}
\end{equation}
where $n \in {\mathbb Z}$.  Thus, the gauge transformation on $A_5$ 
is given by
\begin{equation}
    A_5(x^\mu, y) \rightarrow A_5(x^\mu, y) + \frac{2n}{R},
\label{eq:A5-su5}
\end{equation}
that is, we can make the transformation, Eq.~(\ref{eq:A5-su5}), without 
changing the boundary conditions of the fields.  This means that the 
physically inequivalent vacuum configuration is parameterized as 
\begin{equation}
    A_5 \equiv \langle A_5(x^\mu, y) \rangle = - \frac{a}{R},
\label{eq:a}
\end{equation}
where $0 \leq a < 2$.

Under the orbifold parity $y \rightarrow -y$, $A_5$ has an odd 
transformation property.  This quantizes the possible values of $a$ 
to be $0$ or $1$.  The $a = 0$ case is obviously allowed. In the case 
of $a = 1$, $A_5$ changes the sign under $y \rightarrow -y$, but the 
transformed value is gauge equivalent to the original value, so that 
this case is also allowed. In other words, we can redefine the 
``parity'' as the combined transformation of $y \rightarrow -y$ and 
Eq.~(\ref{eq:theta-su5}) such that the vacuum configuration is 
invariant under this ``parity''.  The breaking of $SU(5)$ only occurs 
when $a=1$.  Thus we concentrate on the case $a = 1$ from now.  This 
gives the holonomy
\begin{equation}
    T = \left( \begin{array}{ccccc}
        1 &&&& \\ &1&&& \\ &&1&& \\ &&&-1& \\ &&&&-1
        \end{array} \right),
\label{eq:vac-su5}
\end{equation}
which clearly breaks $SU(5)$ down to $SU(3) \times SU(2) \times 
U(1)$.  Note that here the boundary conditions for the fields under 
$y \rightarrow y + 2\pi R$ are taken to be trivial, i.e. periodic 
boundary conditions, and the above non-trivial holonomy, 
Eq.~(\ref{eq:vac-su5}), comes from the non-vanishing background gauge 
field, $\langle A_5 \rangle \neq 0$.

Let us now work out the wavefunctions for various fields. Under the 
vacuum configuration Eq.~(\ref{eq:vac-su5}), the ``parity'' is defined 
by the combined transformation of $y \rightarrow -y$ and the large gauge 
transformation given by Eq.~(\ref{eq:theta-su5}).  Therefore the boundary 
conditions for the fields $\phi_+$ ($\phi_-$) that transformed as even 
(odd) under the $y \rightarrow -y$, such as $A_\mu$ ($A_5$), are now 
given by
\begin{equation}
    \phi_{\pm}(x^\mu, -y) = \pm {\rm e}^{i \frac{2 Y_{\phi}}{R} y}
    \phi_{\pm} (x^\mu, y).
\label{eq:orb-ref}
\end{equation}
The mode expansions which satisfy the above boundary conditions are 
easily found as
\begin{eqnarray}
    \phi_{+}(x^\mu, y) &=& \sum_{n=0}^{\infty} \phi_{+ n}(x^\mu)\,
    {\rm e}^{-i \frac{Y_\phi}{R} y} \cos \frac{n}{R} y
    \qquad\quad\:\:\, {\rm for} \:\: Y_\phi:\, {\rm integer},
\label{eq:wf-su5-1} \\
    \phi_{+}(x^\mu, y) &=& \sum_{n=0}^{\infty} \phi_{+ n}(x^\mu)\,
    {\rm e}^{-i \frac{Y_\phi}{R} y} \cos \frac{n+\frac12}{R} y
    \qquad {\rm for} \:\: Y_\phi:\, {\rm half}\:\: {\rm integer},
\label{eq:wf-su5-2} \\
    \phi_{-}(x^\mu, y) &=& \sum_{n=0}^{\infty} \phi_{- n}(x^\mu)\,
    {\rm e}^{-i \frac{Y_\phi}{R} y} \sin \frac{n+1}{R} y
    \qquad\, {\rm for} \:\: Y_\phi:\, {\rm integer},
\label{eq:wf-su5-3} \\
    \phi_{-}(x^\mu, y) &=& \sum_{n=0}^{\infty} \phi_{- n}(x^\mu)\,
    {\rm e}^{-i \frac{Y_\phi}{R} y} \sin \frac{n+\frac12}{R} y
    \qquad {\rm for} \:\: Y_\phi:\, {\rm half}\:\: {\rm integer}.
\label{eq:wf-su5-4}
\end{eqnarray}
Substituting these expansions into the original five-dimensional 
action, we obtain the KK mass spectrum $m_{n} = n/R$, 
$m_{n} = (n+1/2)/R$, $m_{n} = (n+1)/R$ and $m_{n} = (n+1/2)/R$
for each of the above four cases. Note that the phase factor ensures 
that all modes are periodic under $y \rightarrow y + 2 \pi R$.

Here we stress an important difference between the gauge breaking on 
orbifolds and on manifolds.  We have found that we can choose the 
``gauge'' where the boundary conditions under $y \rightarrow y + 2\pi R$ 
are trivial for all the fields.  However, this does not mean that 
all gauge breaking effects are removed entirely from the boundary 
conditions, in contrast with the case of manifolds.  In fact, the 
orbifold reflection $y \rightarrow -y$ still carries gauge breaking 
as explicitly seen in Eq.~(\ref{eq:orb-ref}).  In general, in the case 
of gauge symmetry breaking on orbifolds, we cannot go to the basis 
where boundary conditions are completely trivial and all the gauge 
breaking effects come from the background gauge fields (Wilson lines). 
The boundary conditions intrinsically carry gauge breaking which cannot 
be removed by any gauge transformations, although we can exchange the 
location of the gauge-breaking boundary conditions between $T$ and 
$P$ accompanied by the generation of the background gauge field. 
(Of course, we can choose a gauge basis somewhere between these two 
extreme cases: the boundary conditions under both $P$ and $T$ are 
non-trivial and there is a background gauge field with the value 
$a \neq 0$ or $1$.)  The fact that there is no gauge basis where all 
boundary conditions are trivial in the orbifold breaking theory is 
related to the phenomenon of explicit symmetry breaking defects on the 
fixed points discussed in the next section.

The KK spectrum of the gauge fields is easily identified.
For the unbroken gauge generators of $SU(3) \times SU(2) \times U(1)$, 
the gauge fields do not have hypercharge so that the spectrum is given 
by $m_{n} = n/R$.  On the other hand, for the broken generators 
($X$, $Y$ bosons), their hypercharge is $5/6$ (and hence half integer 
in the normalization here), and the mass spectrum is $m_{n} = (n+1/2)/R$. 
This therefore gives exactly the same mass spectrum worked out in 
\cite{Kawamura:2000ev} using the language of orbifold boundary conditions.

One of the attractive features of this type of model is that 
if the Higgs fields live in the bulk doublet-triplet splitting 
is automatic.  One might naively think that since the triplet 
(doublet) Higgses have integer (half integer) hypercharges, 
the triplet Higgses have zero modes but the doublet ones do not. 
However, this situation can easily be changed by introducing 
a discrete ${\mathbb Z}_{2}$ gauge symmetry acting on the Higgs 
fields (see also \cite{Barbieri:2001dm}). Specifically, by 
introducing a ${\mathbb Z}_{2}$ flavor holonomy in addition to 
the gauge holonomy Eq.~(\ref{eq:vac-su5}), we can interchange the 
wavefunctions for the integer and half integer hypercharge fields 
in Eqs.~(\ref{eq:wf-su5-1} -- \ref{eq:wf-su5-4}).  Therefore, we have 
massless Higgs doublets in this case.  From a single Higgs hypermultiplet 
in ${\bf 5}$ representation, we have a freedom to assign even parity 
to ${\bf 5}$ and odd parity to ${\bf 5}^{*}$ chiral multiplets or 
vice versa.  For the first choice, we find the mass spectrum
\begin{equation}
    \begin{array}{c|c|c|c}
        H_{D} & H_{D}^{c} & H_{T} & H_{T}^{c} \\ \hline
        n/R & (n+1)/R & (n+\frac{1}{2})/R & (n+\frac{1}{2})/R
    \end{array}
\end{equation}
where $H_D$ is identified with the up-type Higgs field. The opposite 
parity assignment gives the down-type Higgs field as a massless mode. 
This completes the derivation of the KK mass spectrum in the orbifold 
theory in the (non-conventional) basis where all the fields have 
periodic boundary conditions.

We now discuss the relation between the wavefunctions given in 
Eqs.~(\ref{eq:wf-su5-1} -- \ref{eq:wf-su5-4}) and those given in 
the orbifold boundary condition picture 
\cite{Kawamura:2000ev, Barbieri:2001vh},
which does not involve any phase factor.  The above wavefunctions 
are derived on the constant background of non-zero $A_5$ with trivial 
(non-trivial) boundary conditions under $T$ ($P$).
However, we can gauge away this background by making a gauge 
transformation 
\begin{equation}
    \theta(x^\mu,y) = \frac{y}{R}.
\label{eq:y-lin}
\end{equation}
In this gauge, the phase factors in Eqs.~(\ref{eq:wf-su5-1} -- 
\ref{eq:wf-su5-4}) are removed and the wavefunctions in the orbifold 
boundary condition picture are reproduced. This gauge transformation 
is not periodic; hence, in the picture with $A_5 = 0$, the fields 
have non-trivial boundary conditions under $y \rightarrow y + 2 \pi R$.
The holonomy Eq.~(\ref{eq:vac-su5}) is transformed from a property 
of the background gauge field to a translational boundary condition, 
$y \rightarrow y + 2\pi R$, and the parity transformation now does not 
contain any gauge breaking twist. The two pictures describe the same 
theory and are related by a non-periodic gauge transformation.

Finally, we discuss the relation between the restricted gauge symmetry 
picture of \cite{Hall:2001pg} and what we have done here. We have seen 
that we can transform one theory to another equivalent theory by using 
a non-periodic gauge transformation.  In doing so, we have changed the 
boundary conditions accordingly; if we had not changed the boundary 
conditions for the fields, the non-periodic gauge transformation would 
have led to a different theory.  This reminds us of the situation in 
non-linearly realized spontaneous broken gauge theories. Suppose 
we consider the standard model Lagrangian expanded around the vacuum 
expectation value $v$ of the Higgs field, $h(x) = v + \eta(x)$, 
and consider the low energy effective theory below the scale $v$ with 
the $W$, $Z$ and Higgs bosons integrated out:
\begin{equation}
    {\cal L} = \sum_{X=L,R} \bar{u}_X i \bar{\sigma}^\mu 
      (\partial_\mu - i Q_u A_\mu) u_X 
      + \left( m_u(v) u_L u_R + {\rm h.c.} \right) + \cdots,
\label{eq:sm}
\end{equation}
where $u_{L,R}$ and $A_\mu$ represent the up quark and the photon. 
This Lagrangian still has an ``invariance'' under the $SU(2) \times 
U(1)$ gauge transformation; if we make an $SU(2)_L \times U(1)_Y$ gauge 
transformation together with an appropriate rotation of the quark 
mass, $m_u(v)$, in the $SU(2)_L$ doublet space, we arrive at 
a differently looking, but completely equivalent, theory.  This is 
closely analogous to what we have done here: the above non-linearly 
realized Lagrangian, Eq.~(\ref{eq:sm}), corresponds to our higher 
dimensional theory; $SU(2) \times U(1)$ gauge transformations (but 
not those of $U(1)$ electromagnetism, $U(1)_{EM}$) to non-periodic 
gauge transformations; and the quark mass, $m_u(v)$, to the boundary 
conditions.  More precisely, our higher dimensional theory corresponds 
to the low energy theory like Eq.~(\ref{eq:sm}) obtained by taking 
the limit $v \rightarrow \infty$, $y_u \rightarrow 0$ with $m_u \equiv 
y_u v$ fixed. ($v \rightarrow \infty$ implies that the breaking is 
``explicit''.)

This analogy immediately suggests an important question: what is the 
``unbroken gauge symmetry'' of the system with orbifold gauge symmetry 
breaking?  In the case of the standard model, the unbroken gauge 
symmetry is $U(1)_{EM}$, which is obtained by looking for the 
symmetry that does not transform the Higgs expectation value (quark 
mass). The situation is quite similar in orbifold gauge theories: 
the ``unbroken'' gauge symmetry is the one which preserves the boundary 
conditions for the fields. In the example of the five-dimensional 
$SU(5)$ theory considered here, it is given by a set of $SU(3) \times 
SU(2) \times U(1)$ gauge symmetries whose transformation parameters have 
profiles $\cos[n y/R]$ in the extra dimension and $SU(5)/(SU(3) 
\times SU(2) \times U(1))$ ones with the parameters having 
$\cos[(n+1/2)y/R]$.\footnote{
The symmetries explicitly broken by the compactification are $SU(3) 
\times SU(2) \times U(1)$ with the gauge parameters $\sin[r y/R]$ and 
$\cos[r' y/R]$ and $SU(5)/(SU(3) \times SU(2) \times U(1))$ with 
$\sin[r y/R]$ and $\cos[(r'+1/2)y/R]$ $(r,r' \in {\mathbb R}; r' 
\notin {\mathbb Z})$, of which the non-periodic gauge transformation 
of Eq.~(\ref{eq:y-lin}) is a part.  These symmetries correspond to 
$(SU(2) \times U(1))/U(1)_{EM}$ in our analogy.} This is exactly 
what is called restricted gauge symmetry in \cite{Hall:2001pg}. 
In a sense, by restricting the gauge transformation parameters in this 
way, we are taking a ``unitary gauge'': we (partially) fix the gauge, 
keeping only a part of the underlying full gauge symmetry manifest in 
the compactified (non-linearly realized) theory.  Then, satisfying this 
``unbroken'' restricted gauge symmetry is the only constraint on 
our effective field theory.

\section{Symmetry Breaking on Orbifold Fixed Points}
\label{sec:fixed}

In this section we discuss several issues in the orbifold breaking 
of gauge symmetries.  As an example, we consider the models of 
$SU(5)$ breaking \cite{Kawamura:2000ev, Altarelli:2001qj, Hall:2001pg, 
Hebecker:2001wq, Barbieri:2001yz} where there is a non-trivial 
holonomy for the translation $T$ but not for $P$.

In the setup of orbifold symmetry breaking, there exists a point 
where only a subgroup of $G$ is manifest, and on that point incomplete 
multiplets and/or $G$ non-invariant interactions can be introduced 
\cite{Hall:2001pg}.  This idea has recently been used to solve the 
doublet-triplet splitting problem by putting just the Higgs doublets 
on the fixed point $y=\pi R$ where only $SU(3) \times SU(2) \times U(1)$ 
is manifest \cite{Hebecker:2001wq}.  One should wonder if such an 
``explicit'' breaking of the $SU(5)$ gauge invariance would lead to 
any inconsistencies.  However, below we show that such incomplete 
multiplets pass non-trivial consistency checks, as required in gauge 
theories.

First, we emphasize that these explicit breakings are allowed only on
the fixed point where only a subgroup of $SU(5)$ is manifest.  The
bulk must still be $SU(5)$ symmetric to keep unitarity up to the
cutoff of the theory.  For instance, the gauge coupling constants in
the bulk must be unified into $SU(5)$.  Second, in order for the
theory to make sense without color-triplet Higgs on the brane, there
should not be any transitions between the doublet and color-triplet
Higgses.  Indeed, in any ``gauge'' the $X$ and $Y$ bosons have 
wavefunctions proportional to $\cos[(n+1/2)y/R]$, and hence they vanish 
on the brane and cannot cause transitions.  Note that in the example 
of a manifold with a discrete holonomy ({\it e.g.}\/, ${\mathbb R}P^{2}$), 
there is no point where the wavefunction of all broken gauge bosons 
($Y_{l}^{m}$ with odd $l$) identically vanish.  This possibility is 
special to orbifolds and is related to the fact that we cannot find 
a gauge where all the boundary conditions are trivial and the gauge 
symmetry breaking comes entirely from the background gauge field.

Third, the partial wave unitarity is satisfied up to the cutoff of 
the theory.  Consider the process $H_{D} H_{D}^{*} \rightarrow X X^{*}$
where $X$ is the $X$-boson.  In four-dimensional field theory, this 
process can proceed through the $s$-channel exchange of $U(1)_{Y}$ 
or $SU(2)_{L}$ gauge bosons, and also through the $t$-channel 
color-triplet Higgs exchange.  Each diagram behaves as $g^{2} 
E^{2}/m_{X}^{2}$ at high energies $E \gg m_{X}$ in $J=1$ partial wave, 
which would cause violation of partial wave unitarity.  However, two 
diagrams cancel and one obtains an amplitude of $\sim g^{2} \log 
(E^{2}/m_{X}^{2})$, which remains unitarity up to extremely high 
energies.  In our orbifold setup, however, the color-triplet Higgs 
does not exist and hence only the first diagram exists.  It appears 
that partial wave unitarity is violated at an energy scale $E 
\sim \sqrt{4\pi} m_{X}/g$.  It turns out, however, that not only the 
zero mode but also the first mode of the KK-tower of $U(1)_{Y}$ and 
$SU(2)_{L}$ gauge bosons can be exchanged conserving KK-momentum 
at the triple-gauge-boson vertex. (This KK momentum conservation 
is up to a sign: ($1+1+0$) is possible, for example.) The 
triple-gauge-boson vertex for the zero mode is determined by the 
integral of wavefunctions 
\begin{equation}
    g_{5} \int_{0}^{\pi R}dy 
    \left(\sqrt{\frac{2}{\pi R}} \cos \frac{y}{2R} \right)^{2}
    \sqrt{\frac{1}{\pi R}}
    = g_{5} \sqrt{\frac{1}{\pi R}} = g_{4},
\end{equation}
while that for the first KK mode 
\begin{equation}
    g_{5} \int_{0}^{\pi R}dy 
    \left(\sqrt{\frac{2}{\pi R}} \cos \frac{y}{2R} \right)^{2}
    \sqrt{\frac{2}{\pi R}} \cos \frac{y}{R}
    = g_{5} \sqrt{\frac{1}{2\pi R}} = \frac{g_{4}}{\sqrt{2}}.
\end{equation}
On the other hand, the gauge coupling of the zero mode to the Higgs 
doublet on the brane is 
\begin{equation}
    g_{5} \left. \sqrt{\frac{1}{\pi R}}\right|_{y=\pi R}
    = g_{4},
\end{equation}
while that of the first KK mode is 
\begin{equation}
    g_{5} \left. \sqrt{\frac{2}{\pi R}}\cos \frac{y}{R}\right|_{y=\pi R}
    = -\sqrt{2}\ g_{4}.
\end{equation}
Therefore, the diagram with the first KK mode exchanged in the 
$s$-channel has the coupling factor $(g_{4}/\sqrt{2}) (-\sqrt{2}\
g_{4}) = - g_{4}^{2}$ as opposed to $g_{4}^{2}$ in the diagram with 
the zero mode.  The cancellation between the two diagrams leads to a 
high-energy behavior for the amplitude of $\sim g^{2} (E^{2}/m_{X}^{2})
(m_{KK}^{2}/E^{2}) \sim g^{2}$ which preserves unitarity at high 
energies.  One can also check that amplitudes for any combination of 
KK $X$-bosons in the final state satisfy partial wave unitarity.  
Of course, the multiplicity $ER$ of KK $X$-boson states lead to the 
factor $ER$ in amplitude for the fastest-growing channel, and would 
violate unitarity at $g^{2} ER \sim 4\pi$, or $E \sim 4\pi/(g^{2} R)$, 
as expected in any five-dimensional gauge theories.  The point is that 
the incomplete multiplet on the brane does not lead to any additional 
unitarity violation beyond what is expected in five-dimensional 
gauge theories.

One way of realizing a split multiplet on the brane is through a 
non-local operator using the Wilson lines.  For example, in the 
supersymmetric $SU(5)$ theory, the gauge fields in the 5th dimension 
can be represented as a chiral superfield $\Sigma(y)$ in the adjoint 
representation, and hence one can write down a coupling on the brane
\begin{equation}
    \int d^{2}\theta \bar{H} \left( M - \mu {\cal P} e^{\oint 
    dy\Sigma(y)}\right) H,
\label{eq:split}
\end{equation}
where the Higgs multiplets, $H$ and $\bar{H}$, fill complete $SU(5)$ 
multiplets of ${\bf 5}$ and $\bar{\bf 5}$ and are localized on the 
brane, while the Wilson line is the integral over the 5th dimension. 
By a fine-tuning between $M$ and $\mu$, we can achieve split multiplets
on the brane.  This operator can be used even if the brane does not sit 
at the fixed point.  In this case, the unitarity consideration indeed 
leads to inconsistencies unless we include the effect of the 
color-triplet Higgs.  One also sees the same unitarity violation 
if the brane is in $S^{1}$ rather than an orbifold.  Therefore, the 
incomplete multiplet does not arise from this effective operator, but 
by putting it on the brane ``by hand''. Alternatively, this follows 
because the second term in Eq.~(\ref{eq:split}) is a non-local operator 
and is expected to be exponentially suppressed by the volume of the 
extra dimension.

Note that we have discussed the incomplete multiplet of $SU(5)$ 
as an example, but the same argument applies for more general cases. 
In particular, we can introduce fields on the brane which do not 
come from an $SU(5)$ representation; we can even introduce fields whose 
$U(1)$ charges are not quantized.  This is easily understood because 
a bad high energy behavior of a single diagram is canceled by the 
other diagram exchanging the same gauge boson but of a different KK 
level in the $s$-channel, and thus we do not need any ``$SU(5)$ 
partner'' to keep the unitarity at higher energies.

Having passed many non-trivial consistency checks, we therefore 
say that explicit $SU(5)$ breaking is allowed on the brane while the 
bulk is still $SU(5)$ symmetric.  Clearly, orbifolds allow 
wider possibilities beyond conventional spontaneous breaking of 
symmetries. No matter what ``gauge'' we use, the wavefunctions for 
the $X$ and $Y$ gauge bosons $A^X_\mu (x,y)$ vanish at $y=\pi R$, 
in spite of the differing boundary conditions. The theory is invariant 
only under a restricted set of gauge transformations $\theta^X(x,y)$; 
in particular, $\theta^X(x,y)$ has no zero mode \cite{Hall:2001pg}.

We finally comment on the relation between the $SU(5)$ breaking 
``vacuum'' and the $SU(5)$ preserving ``vacuum'', {\it i.e.}\/ 
$a=1$ and $a=0$ in Eq.~(\ref{eq:a}).  In the $SU(5)$ preserving 
``vacuum'', we clearly cannot introduce incomplete multiplets on the 
brane, since it violates unitarity below the cutoff scale.  Then, 
how do the split multiplets appears when we move from the $SU(5)$ 
preserving to the $SU(5)$ violating ``vacua''?  The answer in the 
effective field theory is that we simply cannot move between two 
``vacua''. The potential barrier between the two ``vacua'' is infinity, 
so that they are completely disconnected: there is no physical 
process which causes tunneling between the two. Therefore within 
the field theory there is no contradiction to have two disconnected 
``vacua'' with different particle content; they are just two different 
theories.  One may wonder if two ``vacua'' may be somehow connected 
in a more fundamental theory.  For instance, vacua with different 
numbers of generations can be connected in string theory. It does not 
cause a contradiction, however, because the singular point that connects 
two vacua with different numbers of generations gives a conformal theory 
with no particle interpretation \cite{Kachru:1997rs}.

\section{Conclusions}
\label{sec:concl}

In this paper we have studied gauge symmetry breaking on orbifolds. 
It is known that non-gauge-invariant boundary conditions for 
non-contractible cycles on a manifold are equivalent to vacuum 
expectation values of gauge fields (Wilson lines) and hence represent 
spontaneous breakdown of gauge symmetry.  We have shown, however, that 
there is in general no choice of gauge in which orbifold boundary 
conditions at fixed points can be attributed to vacuum expectation 
values of gauge fields.  Therefore, gauge symmetry breaking due to 
orbifold boundary conditions must be interpreted as a controlled form 
of explicit breaking of gauge symmetry.  Such an explicit breaking 
allows for exotic possibilities such as incomplete multiplets or 
explicitly gauge-non-invariant terms at fixed points.  Despite the 
explicit nature of gauge symmetry breaking, the theory still maintains 
important properties of gauge theories such as partial wave unitarity.

We demonstrated the above points using an $SU(5)$ model on $S^1/Z_2$.
An $SU(5)$ effective field theory on $S^1/Z_2$ can take two possible 
forms: $SU(5)$ may be broken or unbroken. In the unbroken case the 
Wilson line and orbifold boundary conditions both preserve $SU(5)$ 
in the same gauge, while in the broken case they do not. Working in 
the gauge with periodic boundary conditions, the question of $SU(5)$ 
breaking appears to rest on whether the gauge field $A_5$ has a vacuum 
expectation value or not --- and it is tempting to say that these are 
two vacua of the same effective field theory. However, we have found 
that this is not the case.  In the case of broken $SU(5)$, there are 
non-trivial ($SU(5)$-violating) boundary conditions in the orbifold 
reflection, even in the gauge where $\langle A_5 \rangle \neq 0$ and 
the translational boundary conditions are trivial.  This implies that 
in the broken case we can never go to the basis where all gauge breaking 
effects are attributed solely to the non-zero background gauge field.
Therefore, we have argued that the two effective field theories, with 
$SU(5)$ broken and unbroken, are intrinsically different.  In particular, 
the broken theory can have explicit $SU(5)$ breaking multiplets and 
interactions localized at an orbifold fixed point while the unbroken 
theory cannot. For the broken case the zero mode gauge transformation 
$\theta^X(x)$ is not a symmetry of the theory, while in the unbroken 
case it is. The effective field theory for the broken case simply does 
not have the four-dimensional $SU(5)$ symmetry, and has only a restricted 
set of five-dimensional $SU(5)$ transformations which are exact and 
unbroken \cite{Hall:2001pg}.

Even though the two effective field theories are different, it may 
well be that they are the low energy limit of different vacuum 
choices of a single more fundamental theory at higher energies. It 
could be that in this theory both the background gauge field and the 
symmetry breaking localized on the orbifold fixed point are generated 
spontaneously.

\section*{Acknowledgments}

We thank Martin Schmaltz, David Smith and Neal Weiner for 
useful discussions.
Y.N. thanks the Miller Institute for Basic Research in Science 
for financial support.  This work was supported by 
the Department of Energy under contract DE-AC03-76SF00098 
and the National Science Foundation under contract PHY-95-14797.

\end{document}